\documentclass{mem}
\usepackage{natbib}\usepackage{txfonts}\usepackage{balance}
\usepackage{graphicx}
\usepackage[a4paper]{hyperref}
\idline{0}{0}
\begin{document}
\def\teff{$T\rm_{eff }$}
\def\kms{$\mathrm {km s}^{-1}$}
\def\simbolx{Simbol--X}
\def\sax{{\it BeppoSAX}}

\title{
Unveiling GRB hard X--ray afterglow emission with \simbolx{}
}

   \subtitle{}

\author{L. \,Amati\inst{1}, E. \,Maiorano\inst{1}, E. \,Palazzi\inst{1}, 
R. \,Landi\inst{1}, F. \,Frontera\inst{1,2}, N. \,Masetti\inst{1}, \\
L. \,Nicastro\inst{1}
}

\offprints{amati@iasfbo.inaf.it}

\institute{
INAF -- Istituto di Astrofisica Spaziale e Fisica cosmica di Bologna,
Via P. Gobetti 101,
I-40129 Bologna, Italy
\and
Dipartimento di Fisica, Universita di Ferrara, I-44100 Ferrara, Italy
}

\authorrunning{Amati et al.}

\titlerunning{GRB afterglow with \simbolx{}}

\abstract{Despite the enormous progress occurred in the last 10 years, 
the Gamma-Ray Bursts (GRB) phenomenon is still far to be fully understood. 
One of the most important open issues that have still to be settled is the 
afterglow emission above 10 keV, which is almost completely unexplored. 
This is due to the lack of sensitive enough detectors operating in this 
energy band. The only detection, by the \sax/PDS instrument (15-200 
keV), of hard X-ray emission from a GRB (the very bright GRB\,990123), 
combined with optical and radio observations, seriously challenged the 
standard scenario in which the dominant mechanism is synchrotron radiation 
produced in the shock of a ultra-relativistic fireball with the ISM, 
showing the need of a substantial revision of present models. In this 
respect, thanks to its unprecedented sensitivity in the 10--80 keV energy 
band, \simbolx{}, through follow--up observations of bright GRBs detected and 
localized by GRB dedicated experiments that will fly in the $>$2010 time 
frame, will provide an important breakthrough in the GRB field. 
}
\maketitle{}

\section{Introduction}

Gamma--Ray Bursts (GRBs) are short and intense flashes of low--energy 
gamma--rays coming from random directions in the sky at unpredictable times
and with a rate of $\sim$ 300/year as measured by all--sky detectors in
low Earth orbit. In the last 10 years, observations allowed huge steps forward 
in the comprehension of these phenomena, such as their cosmological distance scale, 
their huge luminosities, their host galaxies, 
the likely association of "long" ($\sim$2--1000 s) GRBs
with the collapse of peculiar massive stars and of "short" ($<$$\sim$2 s) GRBs
with the merging of compact objects (NS---NS, NS--BH). See, e.g.,
\citet{Meszaros06} for a recent review. However, there are still several
open issues, one of the most important being the emission
mechanisms in play and their relative contribution to the total radiation.
Follow--up observations at longer wavelengths (X--ray, optical, 
radio) of GRB fields generally lead to the detection of delayed, 
fading emission (the afterglow). According to the general interpretation, 
the afterglow emission is described reasonably well, in the framework of 
the fireball model \citep{Cavallo78,Meszaros97}, as 
synchrotron emission from accelerated electrons when a relativistic shell 
collides with an external medium, the interstellar medium in our case. In 
this scenario the afterglow spectrum at any given time consists 
generally of a four segments power law. The spectral 
and temporal indices are linked together by relationships that depend on 
the geometry of the fireball expansion \citep{Sari98} and the 
properties of the circum--burst environment (density, distribution). The 
average temporal decaying index,$\sim$1.3, and spectral photon index, $\sim$2.2, 
obtained from 
observations \citep{Depasquale06} give an electron spectral index 
p$\sim$2.2$-$2.5, 
which is indeed typical of shock acceleration.

\begin{figure*}[t!]
\centerline{\includegraphics[clip=true,width=10cm]{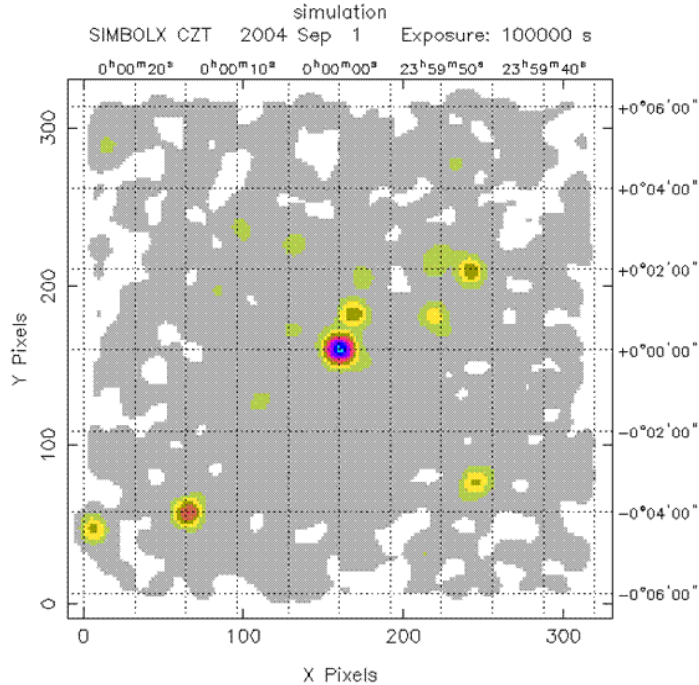}
}
\caption{\footnotesize
Simulated \simbolx{} 15--60 keV image of a bright afterglow (like GRB\,990123) observed for 100 ks starting from 48 hrs since the GRB onset. 
}
\label{image}
\end{figure*}

\section{Afterglow emission in hard X--rays} 

WHile the X--ray afterglow emission of GRB has been widely studied up 
to $\sim$10 keV, 
the upper bound of the energy band of past and presently flying 
X--ray telescopes (e.g., \sax{}, {\it Chandra}, {\it XMM--Newton}, {\it Suzaku}),
in only 
one case, GRB\,990123, the afterglow was detected, by the \sax{} Phoswich 
Detection System (PDS), in the hard X-rays, up to 60 
keV. The multiwavelength 
observations of the afterglow emission (from radio to X--rays) of this event
cannot be 
readily accomodated by basic synchrotron afterglow models \citep{Maiorano05}.
While the temporal and spectral behavior of the optical afterglow 
is possibly explained by a synchrotron cooling frequency between the 
optical and the X-ray energy band, in X--rays this 
assumption only accounts for the slope of the 2--10 keV light curve, but 
not for the flatness of the 0.1--60 keV spectrum. A possible solution 
to the problem was suggested by \citet{Corsi05} including the 
contribution of Inverse Compton (IC) scattering to the hard X-ray emission. 
Anyway, even this IC component is not able to 
provide a self--consistent interpretation of the afterglow. On the other 
hand, leaving unchanged the emission mechanism requires modifying the 
hydrodynamics by invoking an ambient medium whose density rises rapidly 
with radius and by having the shock losing energy. Thus, GRB\,990123
left us 
an open puzzle, showing the need and the importance of afterglow measurements
above 10 keV.

\begin{figure*}[t!]
\centerline{\includegraphics[clip=true,width=8cm,angle=-90]{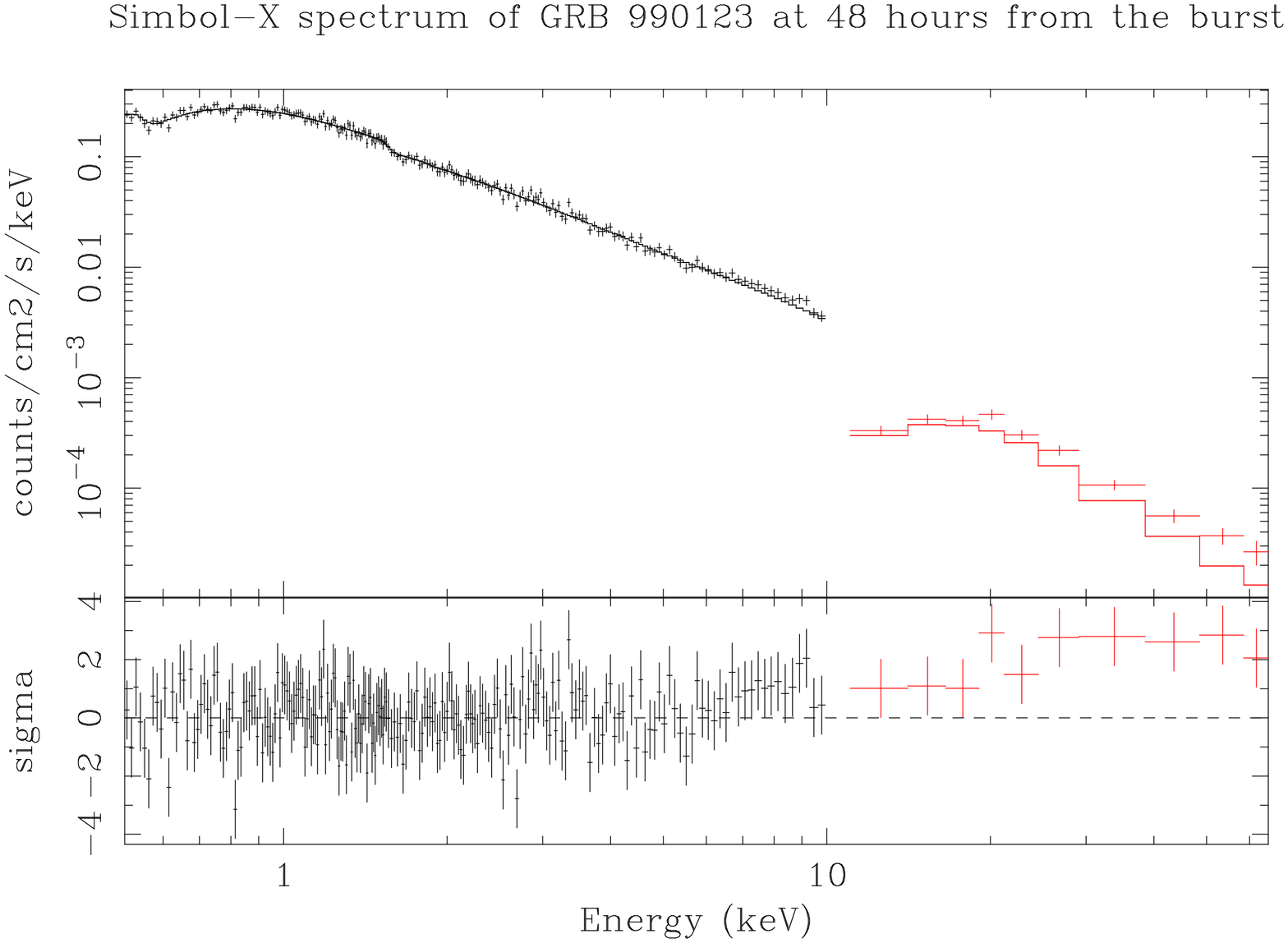}}
\caption{\footnotesize
Simulated \simbolx{} 0.5--60 keV spectrum of the afterglow of GRB\,990123 with an 
observation starting 48 hrs from the GRB onset and an exposure of 100 ks; 
the simulated spectrum (synchrotron + IC component) is fitted with a simple power-law.
}
\label{spectrum}
\end{figure*}

\section{Afterglow measurements with \simbolx{}}

The expected sensitivity of \simbolx{} in the 15--60 keV energy band is of 
the order of 1 $\mu$Crab for a 1 Ms Observation \citep{Ferrando06}, i.e. 
several hundreds times lower than that of any previous instrument 
operating in this energy range. Thus, in principle this mission can open a 
new observational window for the study of GRB afterglow emission and 
provide a big step forward in the comprehension of the physical 
mechanism(s) at play. For instance, at 11hr from the GRB onset about 1/3 
of GRB afterglows show a flux $>$ 100 $\mu$crab \citep{Depasquale06}, 
and thus spectra of very good 
statistical quality up to several tens of keV 
can be obtained with \simbolx{}. A serious concern for 
this kind of studies is the time required to \simbolx{} to be on target 
since a GRB has been detected and localized by other space missions, 
together with the general observational strategy that will be adopted for 
this mission. However, even a 100 ks observation starting about 2 days 
from the GRB onset can provide sensitive spectral measurements in the 
15--60 keV energy band for the bright afterglows. Indeed, the 10\% brightest 
afterglows show a 2--10 keV flux $>$230 $\mu$Crab at 11hr from the burst onset; 
by assuming a typical photon index of 2.1 (i.e. Crab--like) and 
the average temporal decay index (1.3), the expected 15--60 keV flux at 48 
hr is about 35 $\mu$Crab, and the average flux from 48hr to 76 hr 
(corresponding to a 100 ks long observation period) is about 25 $\mu$Crab. 
Fig. 1 shows the simulation of this observation performed with 
the public \simbolx{} tools; as can be seen, the source is well 
detected in the image, with a significance of about 18 $\sigma$ in the 15--60 keV 
energy band. We also show in Fig. 2 the simulation of the 
spectrum of GRB\,990123, generated by assuming a sinchrotron + IC model 
similar to that adopted by Corsi et al. (2005) to fit the  
hard X--ray excess measured by the \sax/PDS, scaled to 48 hr from GRB onset. 
As can be seen, the 
statistical quality of the data allows to characterize the spectrum up to 
about 60 keV and the deviation of the hard X--ray signal (due to the IC 
component) with respect to the simple power-law model (the synchrotron 
component) is clearly detected. The significance of the excess in the 
15--60 keV range is about 6.5 $\sigma$ . We stress that the assumption of an 
observation starting time at 2 days after the GRB is a very conservative 
one, and that pointings with 1 day delay, or even less, could be 
achievable. The simulated image and spectra shown in Fig. 1 and 2 hold also for 
an afterglow of medium intensity pointed at 12 hr from the burst. We also 
remark that, in light of the impact of the only one hard X-ray measurement 
available (GRB\,990123), even a few of such observations can provide a 
significant contribution to the GRB science.

Finally, to do this kind of science, \simbolx{} will need GRB detection and 
location to at least a few arcminutes by other satellites and/or optical 
telescopes. The GRB--related missions operating in the $>$2010 time frame may 
include: Swift (operating since December 2004, GRB detection and arcsec 
localization), SVOM (GRB detection and few arcmin localization), GLAST 
(GRB detection and possibly few arcmin localization), EDGE (GRB detection 
and a few arcsec localization) and possibly GRB detectors on other 
spacecrafts. Thus, also from this point of view GRB follow-up observations 
with \simbolx{} are expected to be feasible.

\section{Conclusions}

Despite the enormous observational progress occurred in the last 10 years, 
the GRB phenomenon is still far to be fully understood.
One of the main open issues is the understanding  of physical mechanisms at the 
basis of prompt and afterglow emission;
the case of GRB\,990123 shows that measurements of the nearly unexplored GRB hard 
($>$ 15 keV) X-ray afterglow emission can provide very stringent test to emission 
models.
Thanks to its unprecedented sensitivity in the 15--60 keV energy band, \simbolx{} 
can provide a significant step forward in this field.
Simulations based on observed distribution of X--ray afterglow fluxes and spectral 
and decay indices show that even a 100 ks TOO observation of a bright GRB 
starting 2 days after the 
event can provide sensitive spectral measurements and allow to discriminate 
different 
emission components for a significant fraction of events.
Moreover, it is likely that significantly lower TOO stat times (12/24 hr) will be 
possible for a few event/year, thus allowing sensitive hard X--rays measurements
also for medium intensity GRB afterglows.
The needed GRB detection and few arcmin localizations will be provided by space 
missions presently planned to be in flight in the $>$2012 time frame and by optical 
telescopes.

\bibliographystyle{aa}

\end{document}